\documentclass[12pt]{article}
\usepackage{latexsym,graphicx,multirow}
\usepackage{amssymb}
\usepackage{amsmath}
\usepackage{amscd}
\usepackage{amsthm}
\usepackage{float}
 \usepackage{setspace}
\usepackage[left=2cm,top=2.5cm,right=2.5cm,bottom=1.5cm]{geometry}
\usepackage{graphicx}
\usepackage{epstopdf}
\usepackage{graphicx,epstopdf}
\usepackage{hyperref}
\DeclareGraphicsExtensions{.eps}
\usepackage{epstopdf} 

\begin{document}
		\begin{center}
		\large{\bf
		 Exploring the  new Tsallis agegraphic dark energy  with interaction through statefinder} \\
		\vspace{10mm}
		\normalsize{Shikha Srivastava$^1$, Umesh Sharma$^2$, Vipin $Dubey^{3\ast}$ }\\
		\vspace{5mm}
		\normalsize
		{$^{1}$ Department of Mathematics,  Aligarh Muslim University,  Aligarh-202002, Uttar Pradesh, India}\\
		{$^{2,3}$ Department of Mathematics, Institute of Applied Sciences and Humanities, GLA University\\
			Mathura-281406, Uttar Pradesh, India}\\

	\end{center}

	\begin{abstract}

	In this work, we explore the recently proposed new Tsallis agegraphic dark energy  model in a flat FLRW Universe by taking the conformal time as IR cutoff with interaction. The deceleration parameter of the interacting new Tsallis agegraphic dark energy model provides the phase transition of the Universe from decelerated to accelerated phase. The  EoS parameter of the model shows a rich behaviour as it can be quintessence-like or phantom-like depending on the interaction ($b^2$) and  parameter $B$. The evolutionary trajectories of the statefinder parameters and  $(\omega_D, \omega_D^{'})$ planes are plotted by considering the initial condition $\Omega_{D}^{0} =0.73$, $H_{0}= 67$ according  to $\Lambda$CDM observational Planck 2018 data  for different  $b^2$ and  $B$. The  model shows both quintessence  and Chaplygin gas behaviour in the statefinder $(r, s)$ and $(r, q)$ pair planes for different  $b^2$ and $B$. 

\end{abstract}



  *Corresponding Author, E-mail: vipin.dubey@gla.ac.in  
\newpage 
	\section{Introduction}
 The surprising discovery of the accelerated expansion
		of the Universe is one of the exciting progress in cosmology
		in the last few years \cite{ref1,ref2,ref3}. However, a search for
		the exact nature of the phenomenon driving this acceleration
		is still under way. In order to explain this behavior,
		two main approches are considered: introducing the
		concept of dark energy (DE) as a new mysterious cosmological
		component or modifying the gravitational part of
		the Einstein equations \cite{ref4,ref5,ref6}. The cosmological constant $\Lambda$ is the simplest approach to DE puzzle, which is responsible for the current acceleration of the Universe expansion and fills
	about $70$  percent of energy content of the cosmos\cite{ref7,ref7a,ref7b,ref8,ref9,ref10}.  It may also be described by modifying general relativity \cite{ref13,ref14,ref14a}.  In addition, recent observations  suggest  a mutual interaction (Q) among the DE and the dark matter (DM) \cite{ref15,ref16,ref17,ref17a,ref18,ref19,ref20,ref20a,ref20b,ref20c},
	 suggesting that their evolution is not independent of one another, which decomposes the total energy-momentum conservation law as

	\begin{eqnarray}
	\label{eq1}
	\rho_{D}^\cdot + 3H\rho_{D}(1+\omega_{{D}}) =-Q,
	\end{eqnarray}
	\begin{eqnarray}
	\label{eq2}
	\rho_{m}^\cdot + 3H\rho_{m} = Q.
	\end{eqnarray}
	 Where $\omega_{D}\equiv \frac{p_{D}}{\rho_{D}}$  is  state parameter of DE. $p_{D}$, $\rho_{m}$ and $\rho_{D}$  are  pressure, DM and DE densities, respectively. Such interaction may solve the coincidence problem \cite{ref21},	
	  and if $Q > 0 (Q < 0)$, then there is an energy transfer from  DE (DM) to DM (DE).  For more details about interacting DE models  \cite{ref16}, can be reviewed .  Despite the fact that the $\Lambda$CDM model is consistent with observations \cite{ref22}, it has some flaws, such as cosmic coincidence and fine-tuning problems. These conditions inspire researchers to look for other dark energy models.\\
	
	An alternative to the $\Lambda$CDM model, in 2007,  based on the uncertainty connection of quantum mechanics, Cai \cite{ref23} proposed the agegraphic dark energy (ADE) model by taking Universe age as IR cutoff. After that a new model of ADE were proposed considering the conformal time as the time scale and  which is not quite the same as the original ADE model \cite{ ref24}.  The authors explored the ADE model in a flat FLRW Universe  with interaction  \cite{ref25}. In \cite{ref26}, the authors implemented the new ADE model with quintessence field. The cosmological and  observational constraints on ADE and NADE models have been explored and suggested that the NADE model has a particular analytic features in the matter-dominated and radiation-dominated era \cite{ref27,ref28}. Similar to the work \cite{ref23,ref24}, recently, a new model of DE has been proposed based on holographic hypothesis and  inspired  by the Tsallis entropy \cite{ref31}, called the Tsallis agegraphic dark energy (TADE) \cite{ref32}. In this work, the authors have also proposed the new Tsallis agegraphic dark energy (NTADE) by taking IR cutoff as conformal time. They studied evolution of cosmos with and without interaction of both models. In papers \cite{ref37,ref38,ref39,ref39aa,ref39aaa}, the authors explored various cosmological parameters and planes of NTADE  in the framework of Chern-Simons modified gravity, Horava-Lifshitz cosmology and  Fractal Cosmology, respectively.
	 While deceleration $(q)$ and Hubble $(H)$ parameters can be used to illustrate the evolution of the Universe, these parameters are unable to distinguish between different dark energy models.
		As a result, the researchers proposed the statefinder pair, which is a nice geometrical analysis \cite{ref40,ref41}.\\

	The author \cite{ref42}, studied the HDE model  by taking IR cut-off as the future event horizon through the statefinder diagnostic. The statefinder diagnostic  is used to discriminate the  two cases of the coupled quintessence scenario by assuming (i) the mass of DM particles relies upon a power law function of the scalar field and in the interim the scalar field evolves in a power law potential  (ii) the mass of DM particles relies exponentially upon the scalar field related to DE and the scalar field evolves in an exponential potential \cite{ref43}. Wei and Cai in \cite{ref44},  studied both the cases   (with and without  interaction)  of ADE models  through statefinder diagnostic and $\omega_{D} - \omega_{D}^{'}$ pair and proposed that the ADE models can easily be  discriminated from the $\Lambda$CDM model. The cosmological evolution of NADE (new agegraphic dark energy) model having interaction among dark  energy and matter  part by utilizing  ${r,s}$ statefinder parameter pair was studied in \cite{ref45}. The authors in \cite{ref46} investigated interacting polytropic gas dark energy model by using  statefinder diagnostic. The reliance of the statefinder parameters on the parameter of the model just as the interaction parameter between dark matter and dark energy is determined. They showed that various estimations of interaction parameter result distinctive evolutionary trajectories in  $\omega_D-\omega_D'$ and $ s - r $ planes.
	 In \cite{ref52,ref53,ref54} researchers discussed the Tsallis holographic dark energy for interacting and non-interacting of flat  and non flat Universe  taking the IR cutoff as Hubble horizon. In the paper \cite{ref55}, the authors investigated the TADE  models without interaction for flat FLRW Universe using the dynamical analysis $\{\omega_D, \omega_D^{'}\}$ pair  and  statefinder analysis,  by considering conformal time as system’s IR cutoffs. 
	  The statefinder, as defined in \cite{ref47,ref48,ref49,ref50,ref51,ref51a}, can effectively distinguish between a wide range of DE models, including holographic Ricci dark energy, ADE, NADE, interacting phantom energy with DM. \\
	 
Inspired by  the above study, an attempt has been made in this  work to analyse the cosmological behaviour of interacting new Tsallis agegraphic dark energy (INTADE) for the flat  FLRW Universe taking the conformal time as system IR cutoff.
	 The statefinder and $\omega_{D}-\omega_{D}^{'}$ pair are also used to  discriminate the INTADE  model from the $\Lambda$CDM. In addition, in consequence of Planck 2018 outcomes VI-$\Lambda$CDM cosmology \cite{ref22}, evolutionary trajectories for the deceleration parameter, EoS parameter,  statefinder and $\omega_{D}-\omega_{D}^{'}$ pair are plotted by considering the present value   of INTADE energy density  ( $\Omega_{D}^{0} =0.73$).  Rest of this paper is organized as: the interacting new Tsallis agegraphic dark energy  and cosmological parameters such as deceleration parameter,  EoS parameter are discussed in Sec. $II$.  To discuss the geometrical behaviour of INTADE model, we obtain statefinder parameters in Sec. $III$. Analysis of the $\omega_{D}-\omega_{D}^{'}$ pair has been discussed in Sec.$IV$. At last in Sect. $V$, we finish up our outcomes.\\
	

	\section{Brief review of the interacting NTADE model}
 The first Friedmann equation of a flat FLRW is written as
 
	\begin{eqnarray}
	\label{eq3}
	H^2= \frac{1}{3m_{p}^2} {(\rho_{m}+\rho_{D})},
	\end{eqnarray}

	where $\rho_{D}$ is INTADE energy density  and $\rho_{m}$ is pressureless matter density. 
	The fractional energy densities is simply instigate as $ \Omega_{n} =\frac{\rho_{n}}{3m_{p}^2H^2}$ for $ n=~ m~$ and $~ D .$ Utilizing this definition, the energy density parameter for matter and INTADE   are $\Omega_{m}=\frac{\rho_{m}}{3m_{p}^2H^2}$ and $
	\Omega_{D}=\frac{\rho_{D}}{3m_{p}^2H^2}$  respectively.
	By putting the energy density parameter for matter and INTADE in  Eq. (\ref{eq2}), we will find  $\Omega_{m}=1-\Omega_{D}$ and the energy densities ratio $\frac{\Omega_{m}}{\Omega_{D}}= r= \frac{1}{\Omega_{{D}}}-1$.\\

 The conformal time is defined as  $dt=ad\eta$  leading $\dot\eta=\frac{1}{ a}$ and  thus
	
	\begin{eqnarray}
	\label{eq4}
	T=\int_{0}^{a} \frac{da}{Ha^2},
	\end{eqnarray}
	\begin{eqnarray}
	\label{eq5}
	\rho_{D}= B \eta^{2\delta-4}.
	\end{eqnarray}
	Where $\eta=\Big({\frac {3{H}^{2} \Omega_{{D}}}{B}} \Big)^{\frac{1}{2\delta-4 }}$
	\begin{eqnarray}
	\label{eq5a}
{\dot \rho_{D}}=\frac {B (2\delta-4)\eta^{2\delta-5}}{{a}}
	\end{eqnarray}

	
	Using the conservation equations given in Eq. (\ref{eq1}) and  Eq. (\ref{eq2}), consolidating the  derivative of Eq. (\ref{eq3}) with time and  using Eq. (\ref{eq5}) and Eq. (\ref{eq5a}),  we get :

	\begin{eqnarray}
	\label{eq6}
	\frac{\dot H}{H^2}=-\frac{3}{2}(1-\Omega_{D})+\frac{(\delta-2)\Omega_{D}} {a \eta H}+\frac{3}{2 b^2}.
	\end{eqnarray}
	By Eq. (\ref{eq6}), the deceleration parameter is  
	\begin{eqnarray}
	\label{eq7}
	q=-1-\frac{\dot H}{H^2}=\frac{1}{2}-\frac{3}{2}\Omega_{{D}}-{\frac {(\delta-2)\Omega_{{D}}}{a \eta H}}-\frac{3}{2 b^2}.
	\end{eqnarray}
	Considering $ Q=3b^2H(\rho_{D}+\rho_{m})$ the mutual interaction between the dark sectors of Universe \cite{ref20}, substituting Eq. (\ref{eq5}) and Eq. (\ref{eq5a}) into Eq. (\ref{eq1}) we find the EoS parameter as:

	\begin{eqnarray}
	\label{eq8}
	\omega_{D}=-1- {\frac { 2\delta-4}{3 a \eta H}}-\frac{b^2}{\Omega_D}.
	\end{eqnarray}
	Using  Eq. (\ref{eq5}) and Eq. (\ref{eq6}),we get the derivatives of the energy density parameter
	
	\begin{eqnarray}
		\label{eq9}
		\Omega_{D}^{'}=\frac {(2\delta-4) \Omega_{{D}}}{a \eta H}+2\Omega_{{D}}  ( 1+q).
	\end{eqnarray}
	Where, prime and dot gives the derivative with respect to $\log a$  and time.\\
	\begin{figure}[htp]
		\begin{center}
				\includegraphics[width=8cm,height=8cm, angle=0]{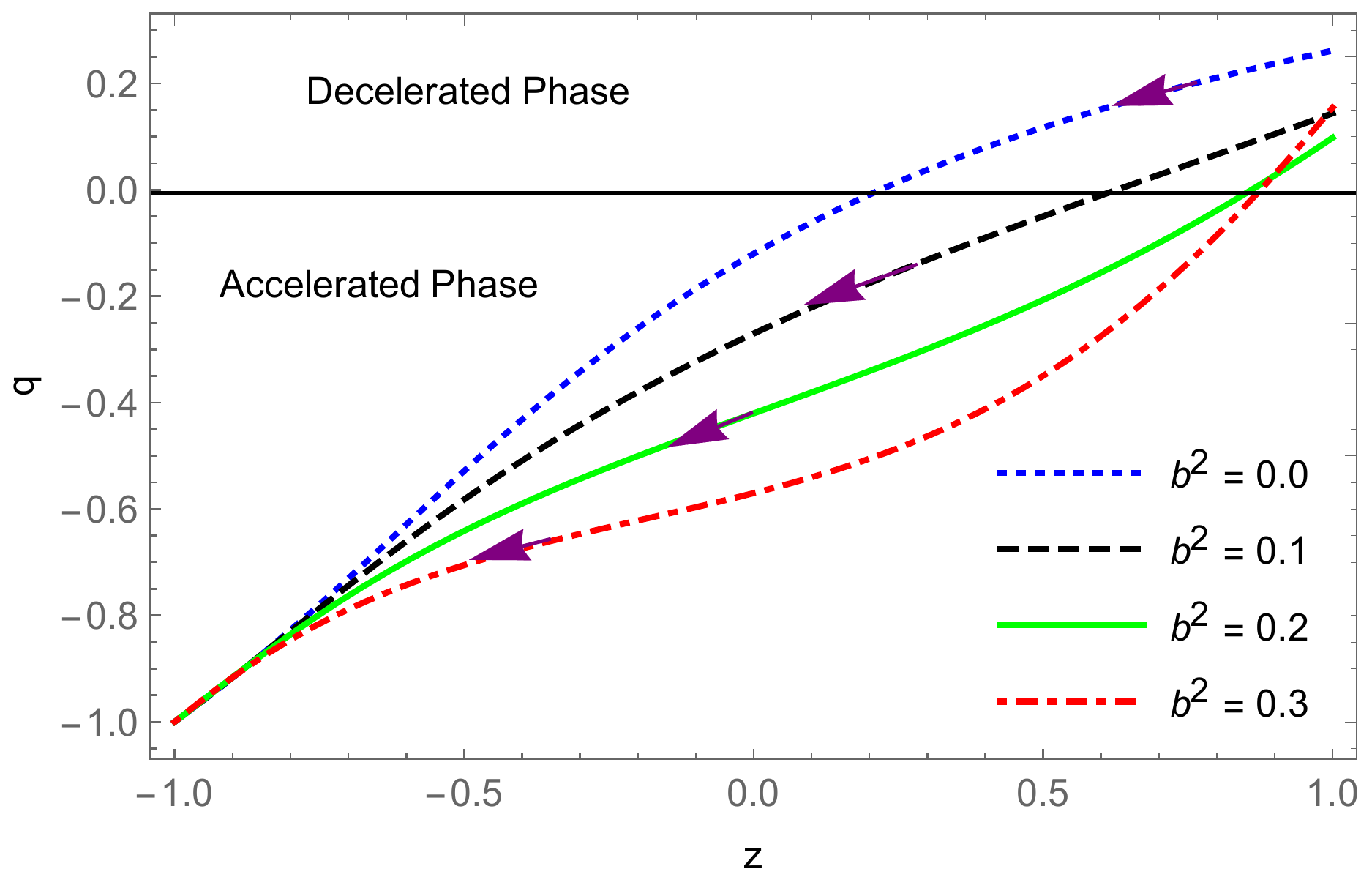}
	\includegraphics[width=8cm,height=8cm, angle=0]{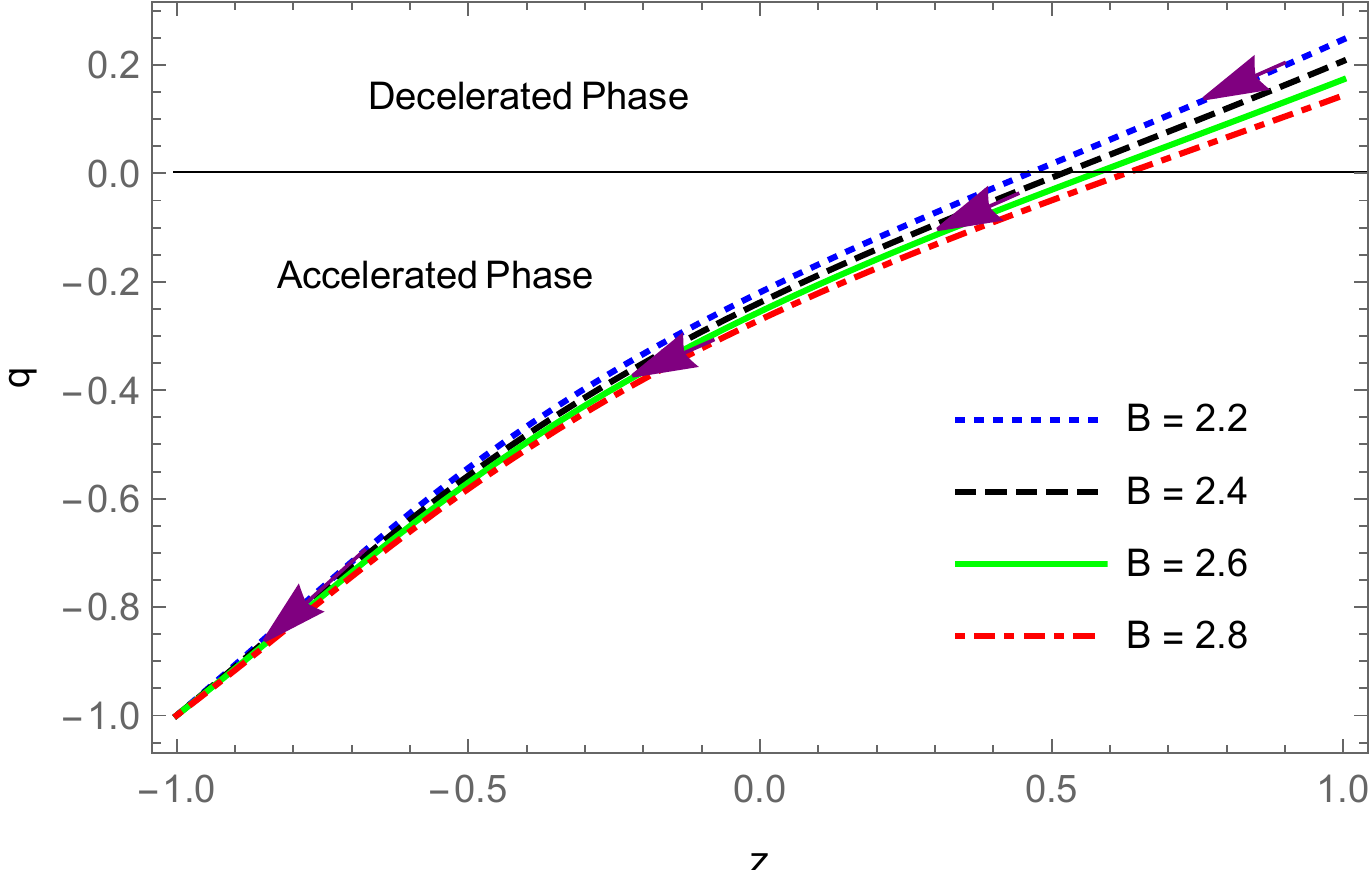}
	
		\caption {Evolution of $q$ (deceleration parameter) versus redshift $z$ for INTADE with initial conditions $\delta = 0.9$, $H(z=0)=67$,  $\Omega_{D}^{0} =0.73$, by considering $B=3$ and different values of coupling  $b^2$ (left panel) and different values of $B$ (right panel) fixing $b^2 = 0.1$. }
		
			\label{fig1}
		\end{center}
	\end{figure}
	
	\begin{figure}[htp]
		\begin{center}
				\includegraphics[width=8cm,height=8cm, angle=0]{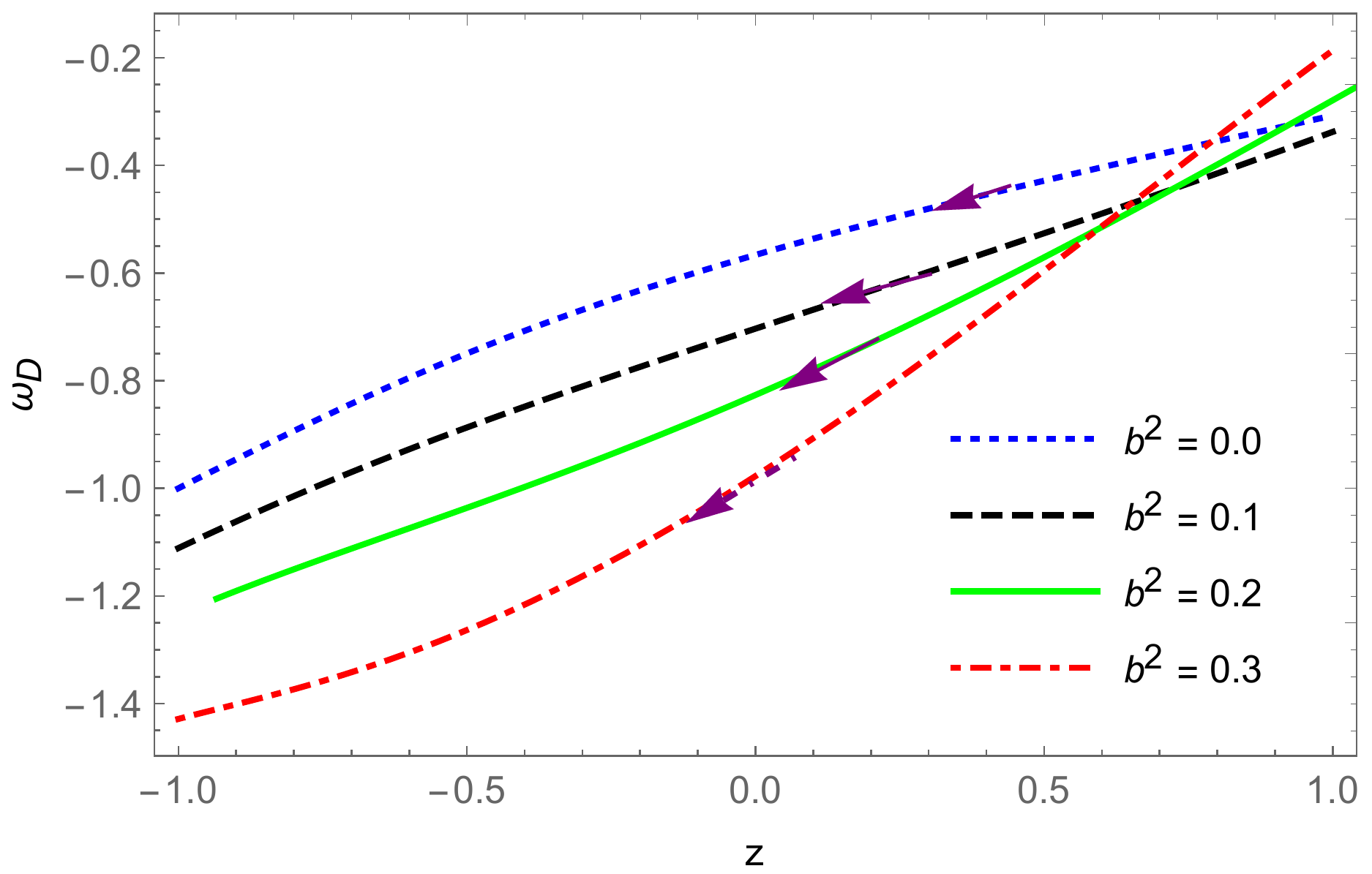}
			\includegraphics[width=8cm,height=8cm, angle=0]{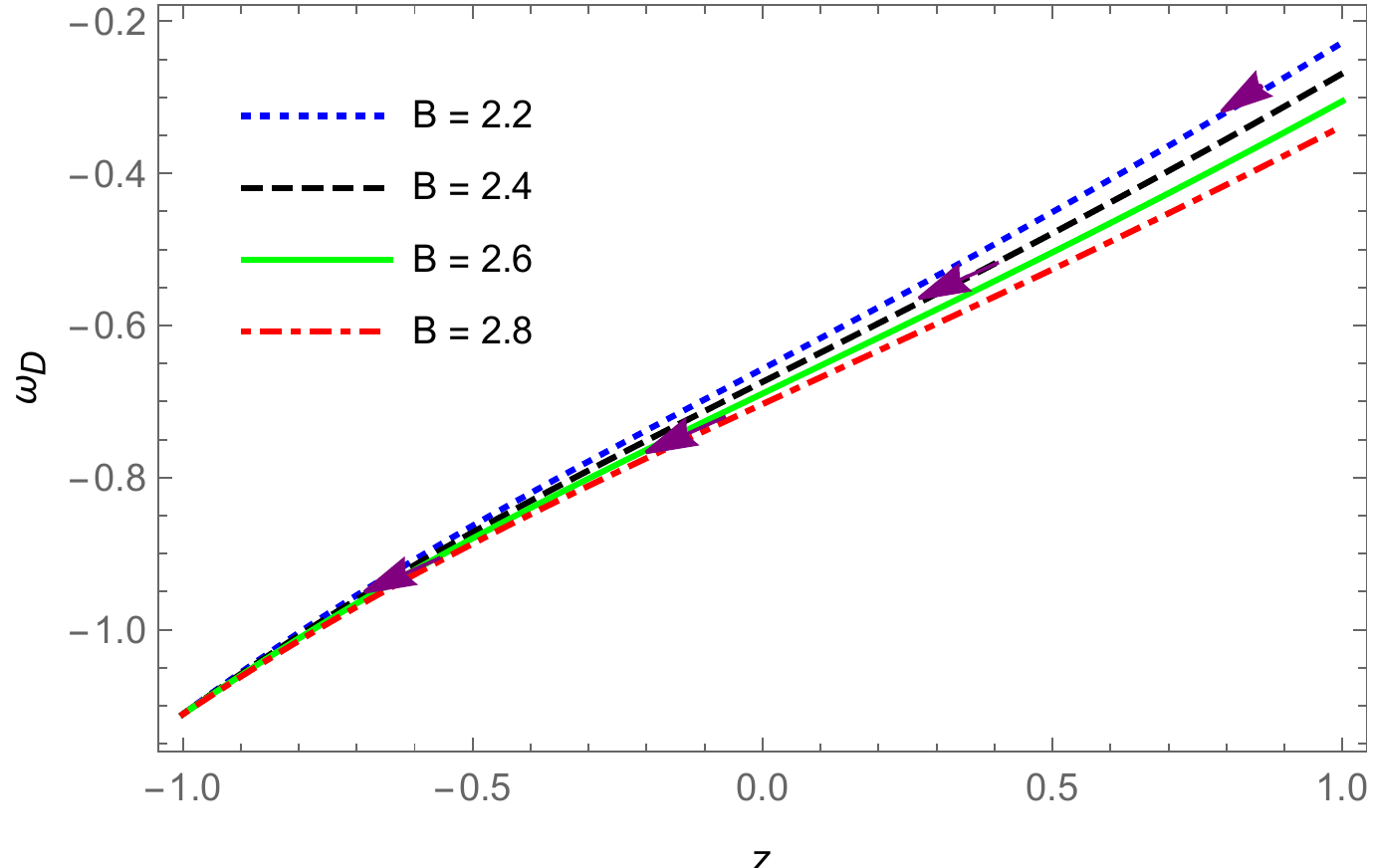}
			\caption{Evolution of $\omega_D$ against  $z$ (redshift parameter) for INTADE with initial conditions $\delta = 0.9$, $H(z=0)=67$,  $\Omega_{D}^{0} =0.73$, by considering $B=3$ and different values of coupling  $b^2$ (left panel) and different values of $B$ (right panel) fixing $b^2 = 0.1$.}
			\label{fig2}
			
		\end{center}
	\end{figure}
	The evolutionary behaviour of the deceleration parameter and EoS parameter is plotted for the INTADE model versus redshift $z$ by finding its numerical solution using the initial values of  $\Omega_D$, as   $\Omega_{D}^{0} =0.73$, and $H_{0}= 67$.  The deceleration or acceleration of the Universe is described by the deceleration parameter $q$. The Universe is in accelerating or decelerating phase accordingly if $q<0$ or $q >0 $,  respectively.
	From Fig. 1, we observe the behaviour of the deceleration parameter against redshift $z$  for different coupling $b^2$ (left panel) and different values of  $B$ (right panel).  It is clear from both panel of Fig. 1, the deceleration parameter $q$ crosses the boundary $q = 0$ (black solid line) from $  q >0$ to $q< 0$. This implies that the universe undergoes decelerated expansion at the early time and later starts accelerated expansion. The transition from decelerated expansion to the accelerated expansion occurs gradually for different interaction ( $b^2$) and different $B$. 
	However, the difference between them is well distinguished, the	transition point increases as interaction $b^2$ grows.

 The EoS parameter $\omega_{D}$ with respect to $z$ is plotted for $\delta=0.9$,   different coupling $b^2$ (left panel) and different values of  $B$ (right panel) in Fig. 2, which depicts that  $\omega_{D}$ evolves from the quintessence region  $-1 < \omega_{D} < -1/3$, at high redshift region for each value of  $b^2$.  At low redshift region for the non-interacting case  $b^2 = 0$,  $\omega_{D}$ approaches the cosmological constant  $\omega_{D} = -1$.  But for interacting cases, $\omega_{D}$ crosses the phantom divide line ($\omega =-1$) and lies in the  phantom region ($\omega <-1$) at low redshift region. The similar behaviour of the deceleration and the EoS parameter has also been observed in  \cite{ref32}.\\

	\section{Statefinder Analysis of Interacting NTADE}
	In this section, we discuss  the INTADE  model through the statefinder analysis.The statefinder indicative pair\cite{ref40,ref41} is given as:
	
	\begin{eqnarray}
	\label{eq10}
	r = \frac{\dddot {a}}{aH^{3}},
	\end{eqnarray}
	\begin{eqnarray}
	\label{eq11}
	s = \frac{r-1}{3 (q - \frac{1}{2})}.
	\end{eqnarray}

	The expression of statefinder parameters $r$ and $s$ in terms of EoS parameter and energy density can be written as:

	\begin{eqnarray}
	\label{eq12}
	r=1+\frac{9}{2}\Omega_{{D}}~\omega_{{D}} (1+\omega_{{D}})-\frac{3}{2}\omega_{D}^{'}~\Omega_{{D}}+\frac{9 b^2}{2}\omega_{{D}},
	\end{eqnarray}
	\begin{eqnarray}
	\label{eq13}
	s=(1+\omega_{{D}})-{\frac {\omega_{D}^{'}}{3\omega_{{D}}}}+\frac{b^2}{\Omega_{{D}}}.
	\end{eqnarray}
	The different DE models can be discriminated from each other and from the $\Lambda$CDM model using  first statefinder parameter $r$,  second statefinder  parameter $s$, ($ r, s$) plane and ($ r, q$) plane. The several dark energy models show distinct evolutionary trajectories of their evolution in  ($ r, s$) and ($ r, q$) plane. Now, we use $r$ (first statefinder  parameter), $s$ (second statefinder  parameter), ($ r, s$) plane and ($ r, q$) plane to discriminate the INTADE model for different values of interaction  $b^2$  and different values of $B$, while fixing other parameters according
	to the best-fit observational values. Also, an important purpose of any diagnostic is that it allows us to discriminate between a given DE model and the simplest of all models -  $\Lambda$CDM. This is exactly done by the  statefinder.  The value of the first statefinder parameter $r$ remains pegged at unity i.e. $r= 1$ for the $\Lambda$CDM model, even as the matter density evolves from a large initial value to a small late-time value. It is easy to show that $({r, s})=({1, 0})$ is  fixed point for $\Lambda$CDM \cite{ref40,ref41}. \\
	\begin{figure}[htp]
		\begin{center}
	
				\includegraphics[width=8cm,height=8cm, angle=0]{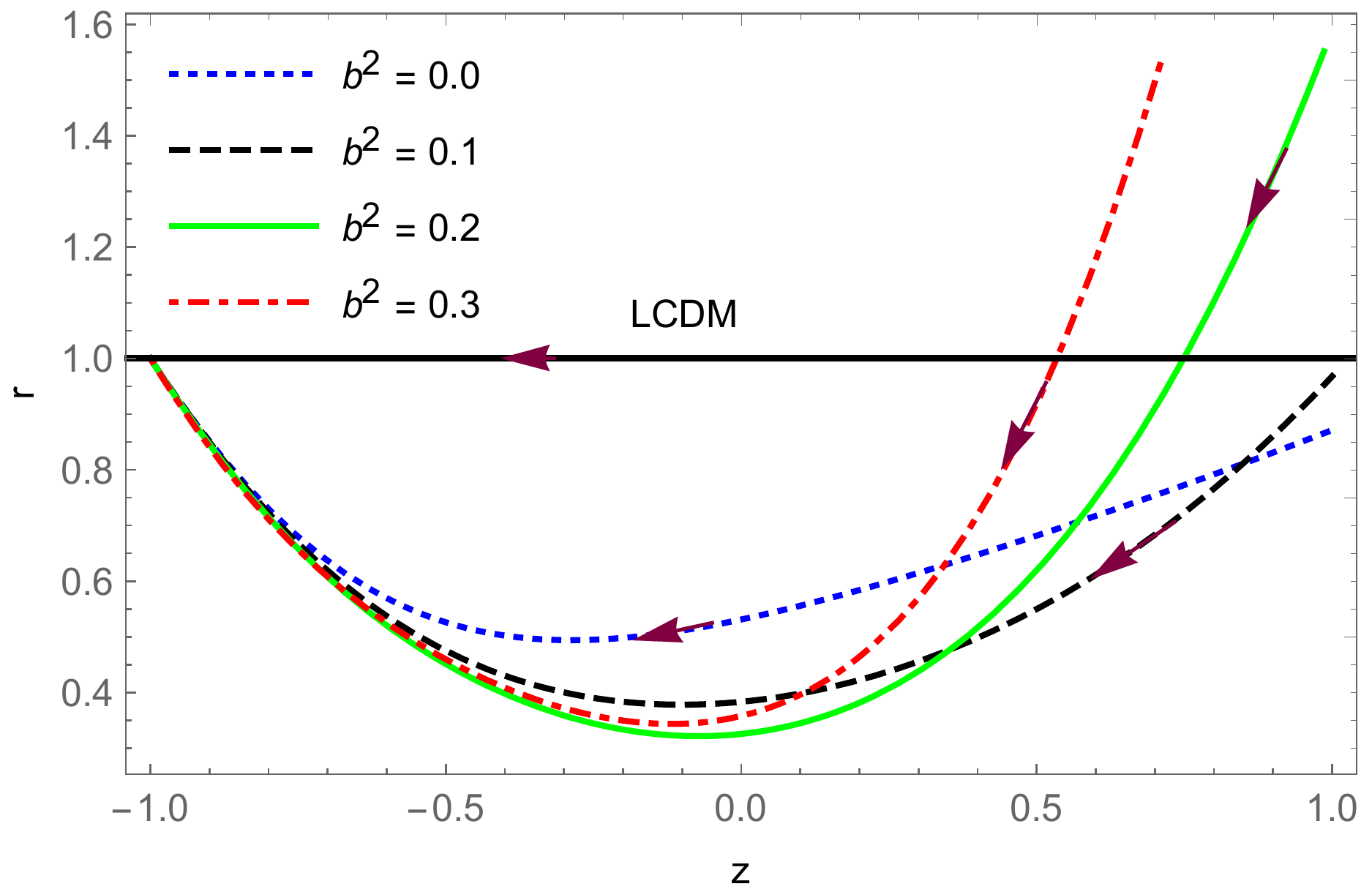}
					\includegraphics[width=8cm,height=8cm, angle=0]{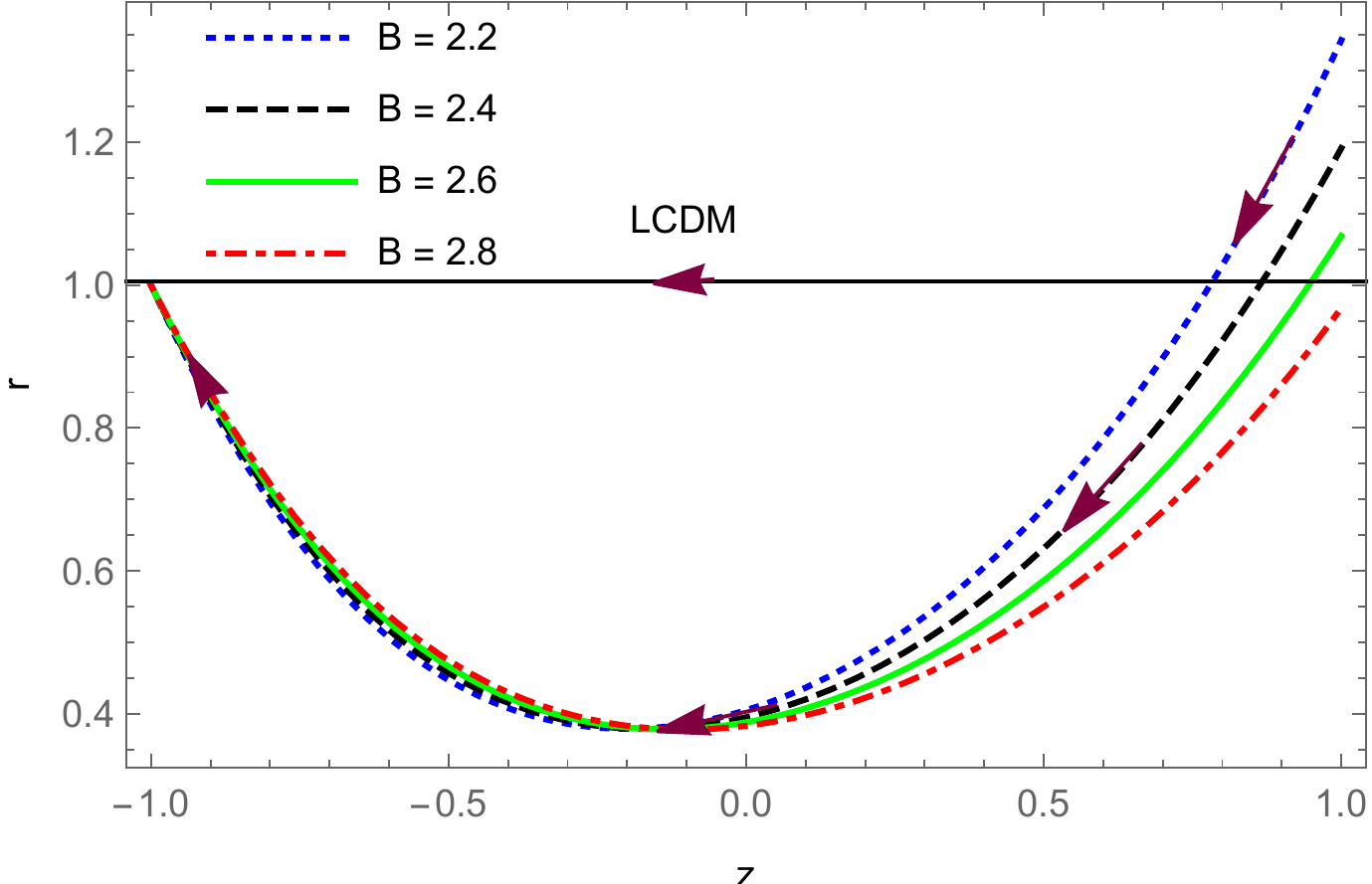}
			\caption{Evolution of $r$ versus $z$ for INTADE with initial conditions $\delta = 0.9$, $H(z=0)=67$,   $\Omega_{D}^{0} =0.73$, by considering $B=3$ and different values of coupling  $b^2$ (left panel) and different values of $B$ (right panel) fixing $b^2 = 0.1$.}
			\label{fig3}
		\end{center}
	\end{figure}
	\begin{figure}[htp]
		\begin{center}
	
				\includegraphics[width=8cm,height=8cm, angle=0]{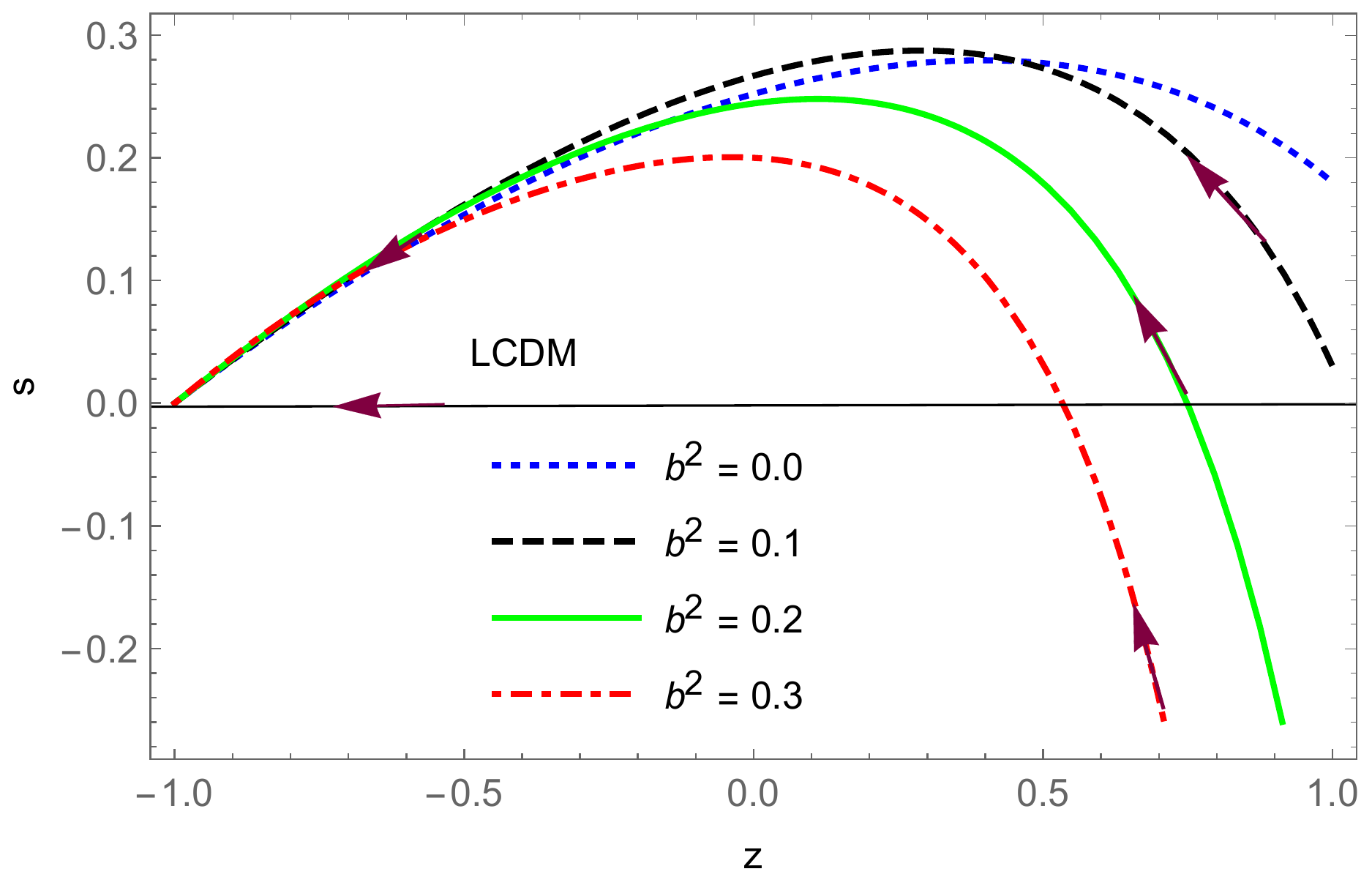}
					\includegraphics[width=8cm,height=8cm, angle=0]{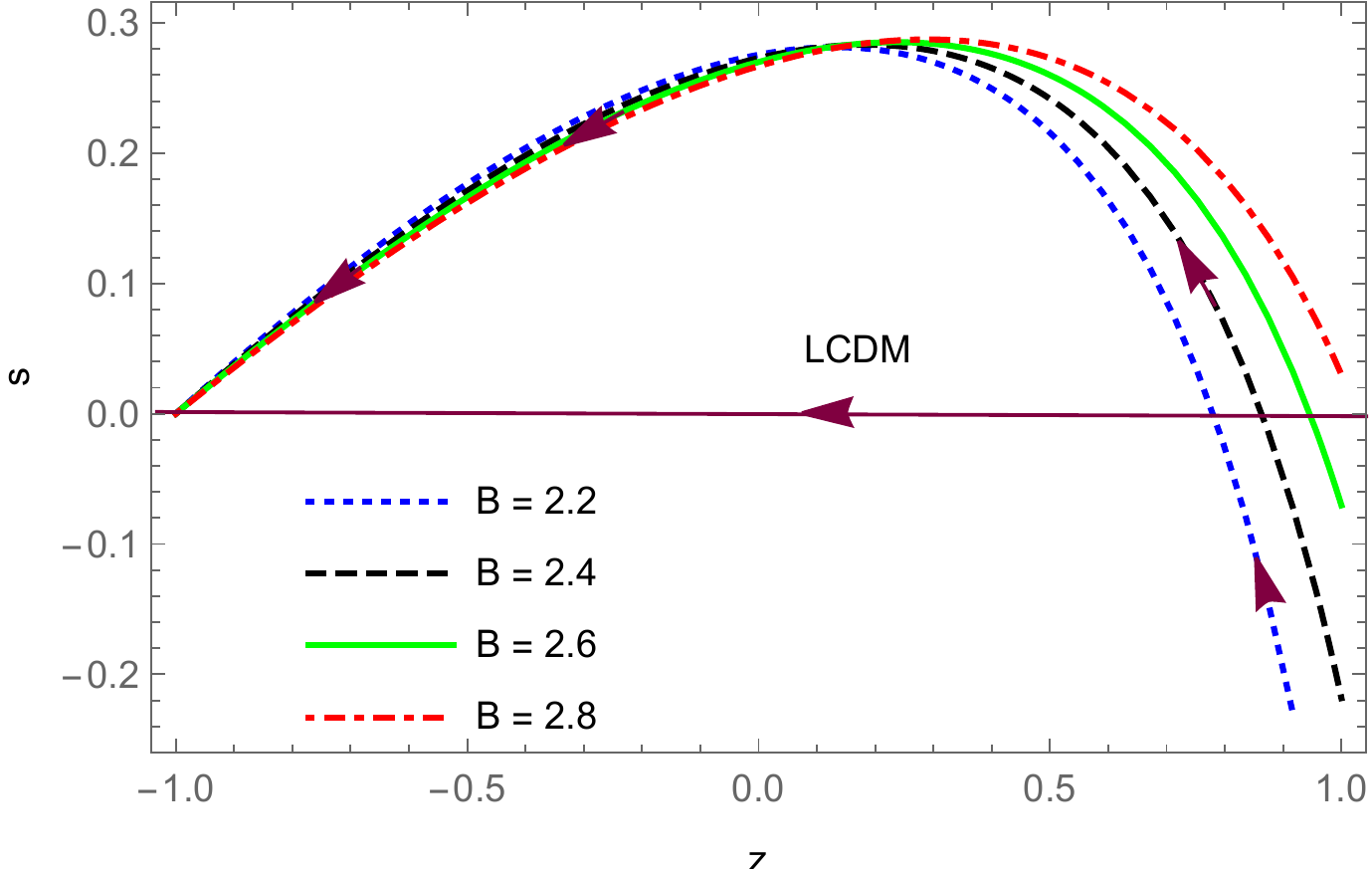}
			\caption{Evolution of $s$ (statefinder parameter) versus $z$ (redshift parameter) for INTADE with initial conditions $\delta = 0.9$, $H(z=0)=67$,  $\Omega_{D}^{0} =0.73$, by considering $B=3$ and different values of coupling  $b^2$ (left panel) and different values of $B$ (right panel) fixing $b^2 = 0.1$.}
			\label{fig4}
		\end{center}
	\end{figure}
	\begin{figure}[htp]
		\begin{center}

				\includegraphics[width=8cm,height=8cm, angle=0]{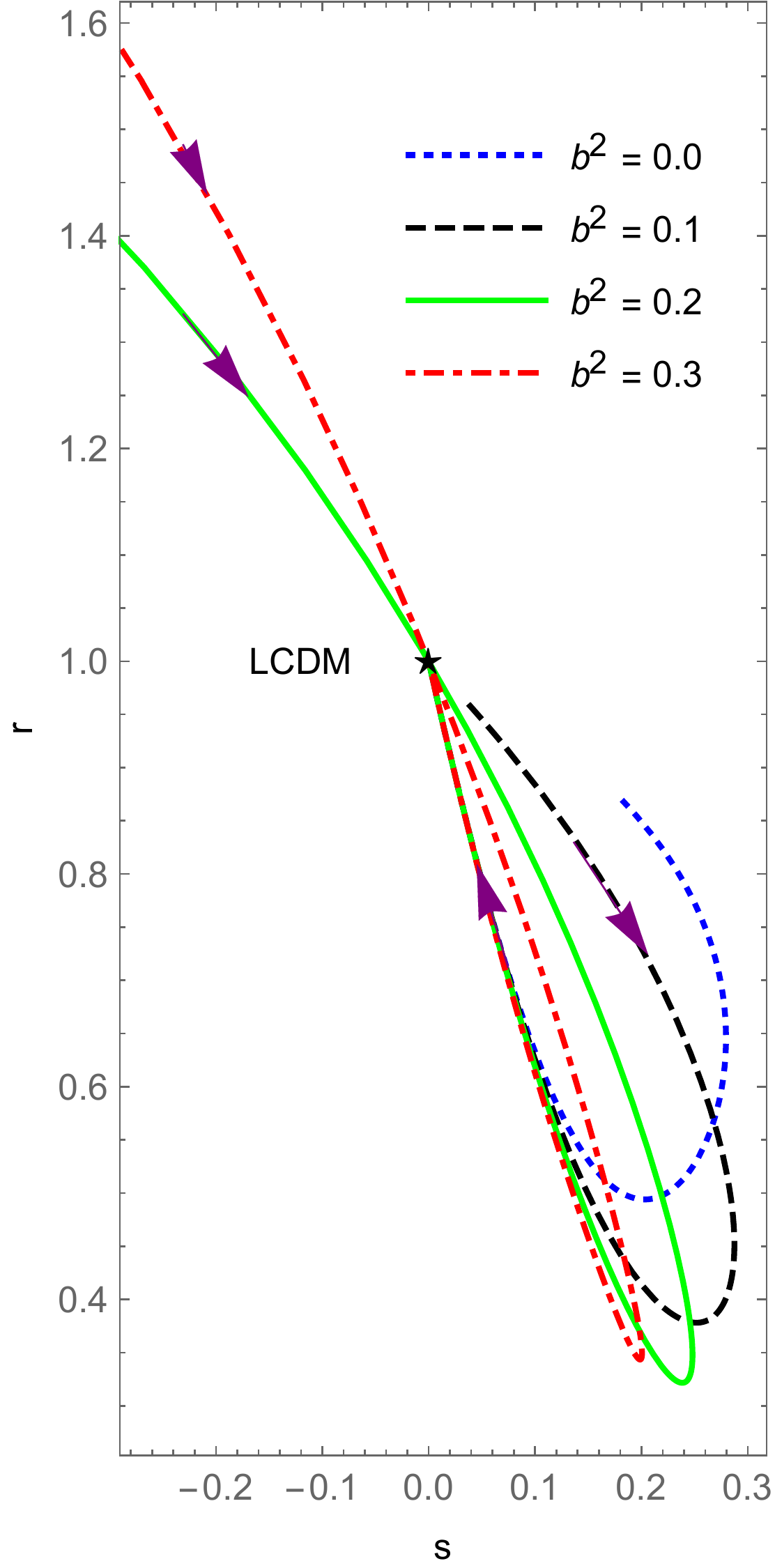}
					\includegraphics[width=8cm,height=8cm, angle=0]{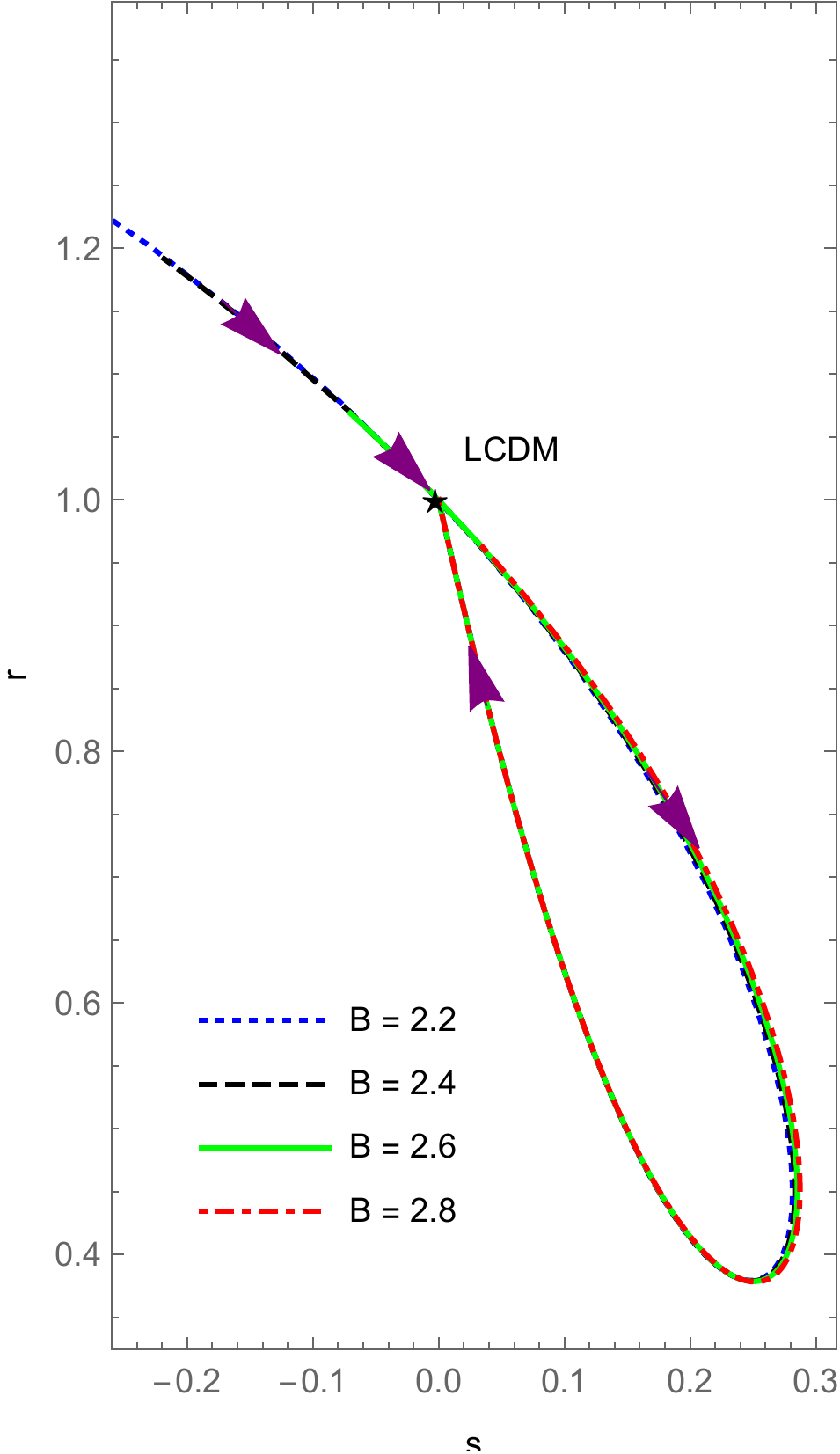}
			\caption{Evolution trajectories of the statefinder in the $r-s$ for  INTADE with initial conditions $\delta = 0.9$, $H(z=0)=67$,  $\Omega_{D}^{0} =0.73$,  by considering $B=3$ and different values of coupling  $b^2$ (left panel) and different values of $B$ (right panel) fixing $b^2 = 0.1$.}
			\label{ fig5}
		\end{center}
	\end{figure}
	\begin{figure}[htp]
		\begin{center}
	
				\includegraphics[width=8cm,height=8cm, angle=0]{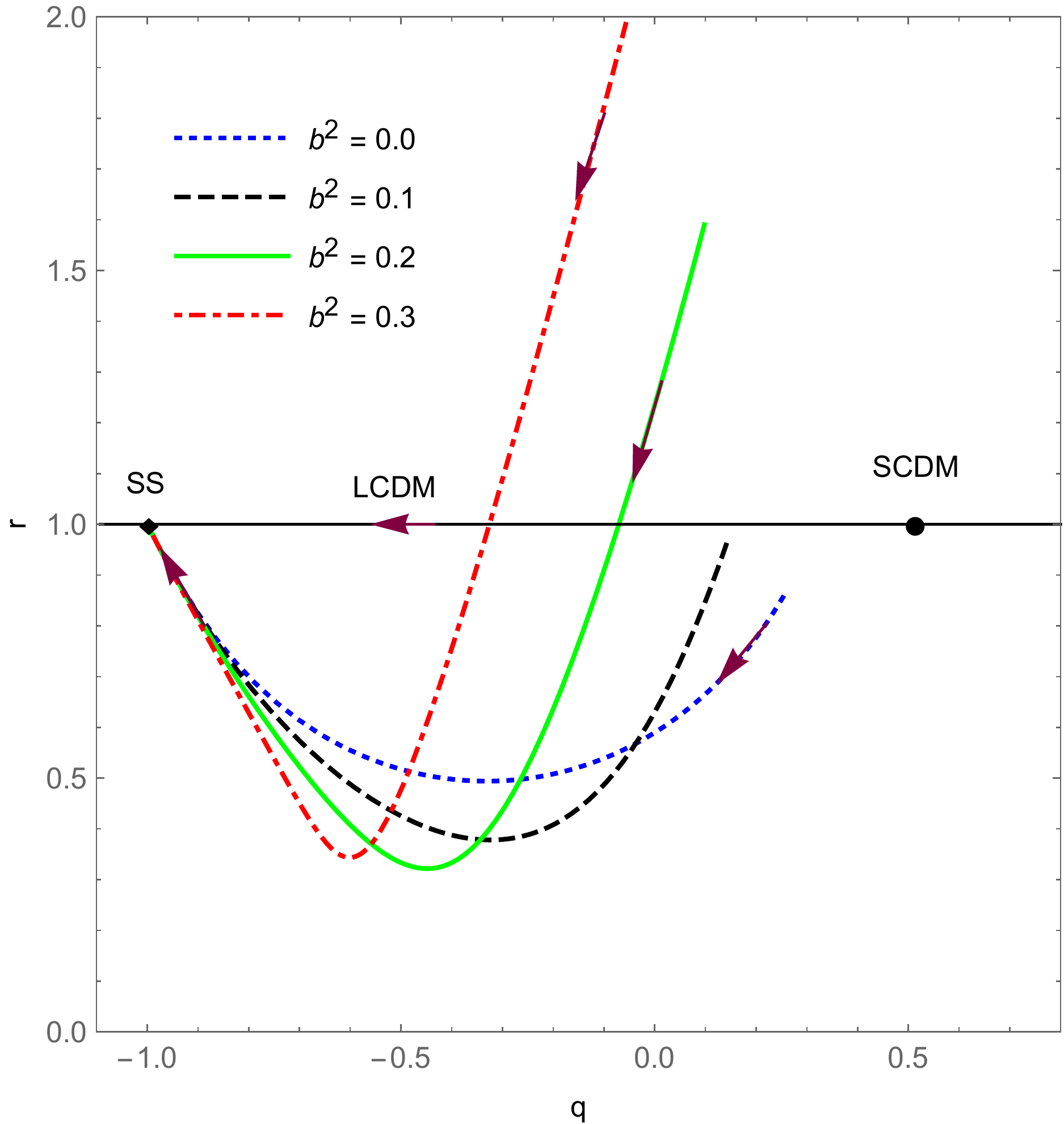}
					\includegraphics[width=8cm,height=8cm, angle=0]{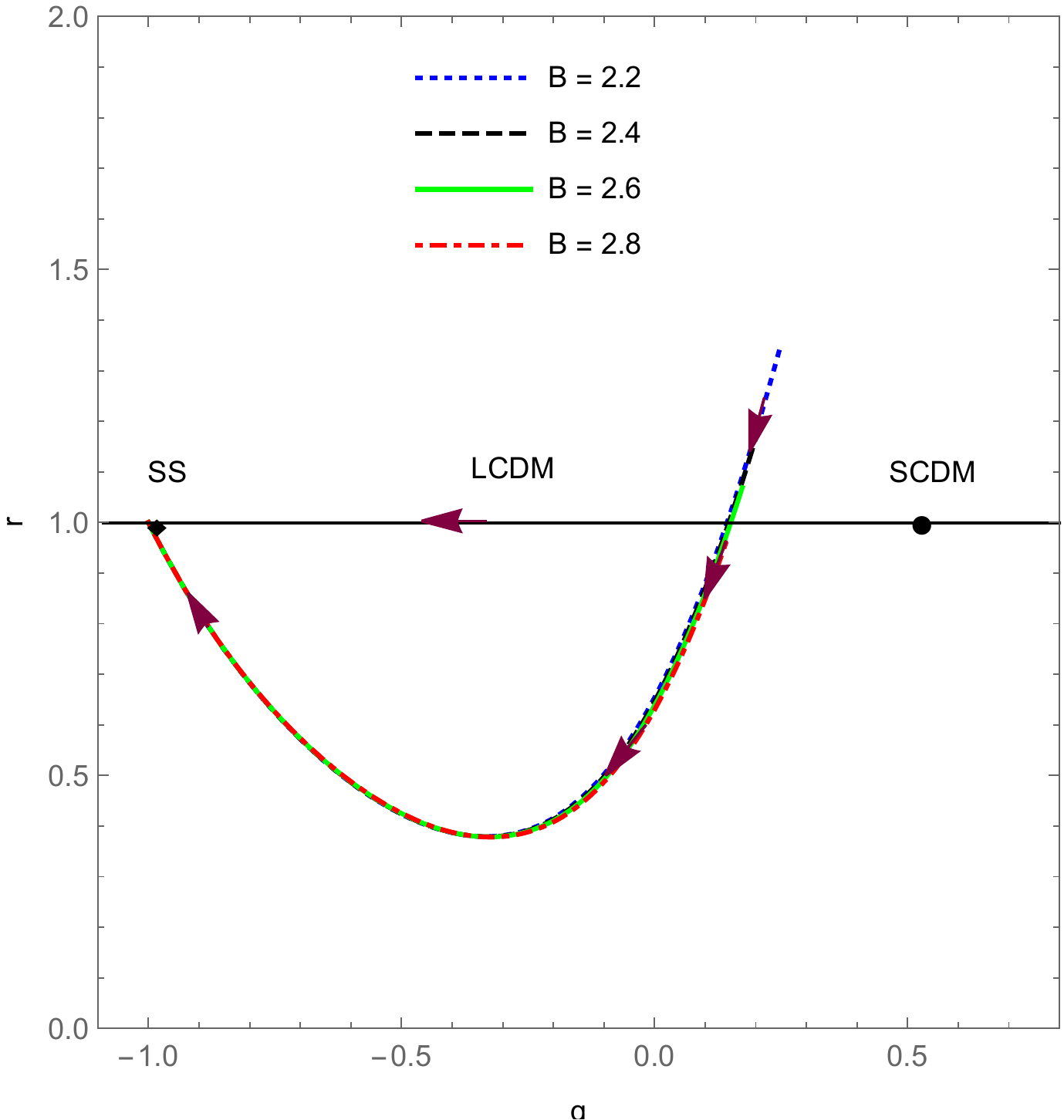}
			
			\caption{Evolution of  $r$ ( first statefinder parameter)  verses $q$ (deceleration parameter) for INTADE with initial conditions $\delta = 0.9$, $H(z=0)=67$,  $\Omega_{D}^{0} =0.73$, by considering $B=3$ and different values of coupling  $b^2$ (left panel) and different values of $B$ (right panel) fixing $b^2 = 0.1$.}	
			\label{fig6}
		\end{center}
	\end{figure}
	
	We have plotted the first statefinder parameter $r$ against redshift $z$ in Fig. 3  for different coupling $b^2$ (left panel) and different values of  $B$ (right panel). It can be seen from Fig. 3, $r$ evolves below the $\Lambda$CDM line $r = 1$ for  $b^2 = 0$,  $b^2 = 0.1$ and  above the $\Lambda$CDM line $r = 1$  for $b^2 = 0.2$,  $b^2 = 0.3$. The curve of $r(z)$ first decreases from past to present and then increases monotonically, finally approaches to the $\Lambda$CDM at future. The evolutionary trajectories can be well discriminated from the  $\Lambda$CDM at high redshift.  

	From Fig. 4, we observe the behaviour of the second statefinder parameter $s$ versus redshift $z$  for different coupling $b^2$ (left panel) and different values of  $B$ (right panel). It can be seen that the evolutionary trajectories are more distinguishable at high redshift region and approach to $s=0$ in the future. We observe from Fig. 3 and 4, that both the statefinder parameters, $r$ and $s$ deviate significantly from the $\Lambda$CDM at high redshift region and both reaches to $\Lambda$CDM in the future.\\

	 Alam et al. \cite{ref41} presented the evolutionary behaviour of the statefinder pair
	$(r, s)$. They have suggested  that the vertical line at $s = 0$ effectively divides the
	$r–s$ plane into two halves. The left half contains Chaplygin gas (CG)
	models that commence their evolution from  $s = -1$, $r = $1 and end it
	at the LCDM fixed point $(s =0, r = 1)$ in the future. The quintessence
	models occupy the right half of the $r–s$ plane. These models commence their evolution from the right of the LCDM fixed point and,
	such as CG, are also attracted towards the LCDM fixed point in the
	future   \cite{ref41}. The authors in  \cite{refu0}  observed that the $(r, s)$ curves have two
		branches on two sides of the asymptote. The branch on the right hand side of the asymptote corresponds
		to decelerating phase before (or up to) dust era, while the left hand side branch has a transition from
		decelerating phase upto $\Lambda$CDM era ( see Fig. 4 of \cite{refu0}).

	The evolutionary trajectory of the statefinder pair $(s, r)$ for the INTADE model is graphed in Fig. 5  for different coupling $b^2$ (left panel) and different values of  $B$ (right panel). This shows that the $(s, r)$ curve evolves from quintessence region ($r <1$, $s >0$) at an early time for all values of  $b^2$ and approaches to the  $\Lambda$CDM point  ($r = 1$, $s = 0$), which is shown as a star in the figure. But for $b^2 = 0.2$,  $b^2 = 0.3$  INTADE model shows the Chaplygin gas behaviour ($r >1$, $s <0$) at late time.  It is also important to mention here that the interaction ($b^{2}\neq 0$) between matter and dark energy changes the
		evolutive trajectory of $r(s)$.  The  interaction between matter and dark energy
		affects the evolutive process of the Universe, but not the fate
		of the Universe. We can easily see that the trajectory of $r(s)$
		will pass the fixed point $\{r = 1, s = 0\}$ of $\Lambda$CDM in the	future and is similar to some of  dark energy models \cite {refu1,refu2,refu3,refu4,refu4a}.  The statefinder analysis has also been performed for the HDE, THDE, interacting THDE and ADE  models by the authors in \cite{ref42,ref52,ref53,ref44}.
 The HDE, ADE, and THDE models have similar evolutionary behavior to the INTADE model in the $ s - r $ plane, while the NTADE model has different evolutionary behavior in the $ s - r $ plane compared to the interacting THDE model  \cite {ref53}.\\

 In Fig. 6, we have graphed  evolutionary trajectories of the statefinder pair ($q-r$) of  INTADE model  for different coupling $b^2$ (left panel) and different values of  $B$ (right panel). The fixed point $(q = 0.5, r = 1)$ presents the SCDM i.e. the matter dominated and (q = -1, r = 1) presents the SS model i.e. de-Sitter universe, respectively.  The evolutionary trajectory of the ($q-r$) plane of INTADE model starts from the left of a matter-dominated universe i.e. SCDM ( r = 1, q = 0.5) in the past and decreases monotonically and finally reaches the de-Sitter expansion (SS) $(q = -1, r = 1)$ for different values of $b^2$, in the future.  The evolutionary trajectories of the ($q-r$) plain are well separated at low redshift region for different values of interaction  $b^2$.

	\section{Analysis of the $\omega_{D}-\omega_{D}^{'}$ pair}
	
		In this section, we study the $\omega_{D}- \omega_{D}^{'}$ pair dynamical analysis for the NTADE model which is widely used in the previous works. The fixed point $\omega_{D} = -1$, $\omega_{D}^{'}=0$ denotes the standard $\Lambda$CDM  in the $\omega_{D}-\omega_{D}^{'}$ plot. The scopes of the DE quintessence model  have been explored in \cite{ref78} in the $\omega_{D}-\omega_{D}^{'}$ plane. The dynamical property of other dark energy models have been utilised from $\omega_{D}-\omega_{D}^{'}$  viewpoint \cite{ref79,ref80,ref80a,ref80b,ref81}.

	\begin{figure}[htp]
		\begin{center}

				\includegraphics[width=8cm,height=8cm, angle=0]{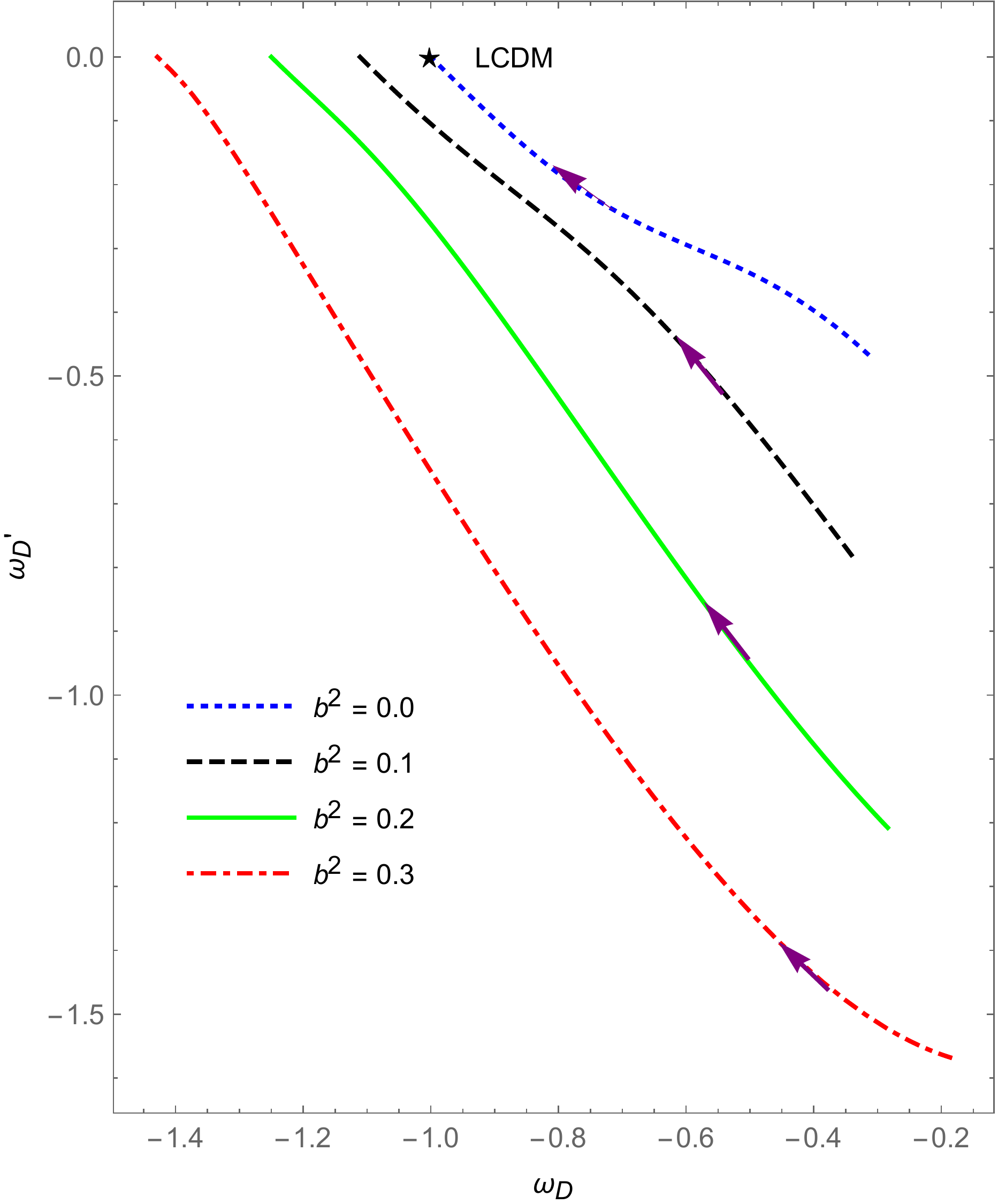}
					\includegraphics[width=8cm,height=8cm, angle=0]{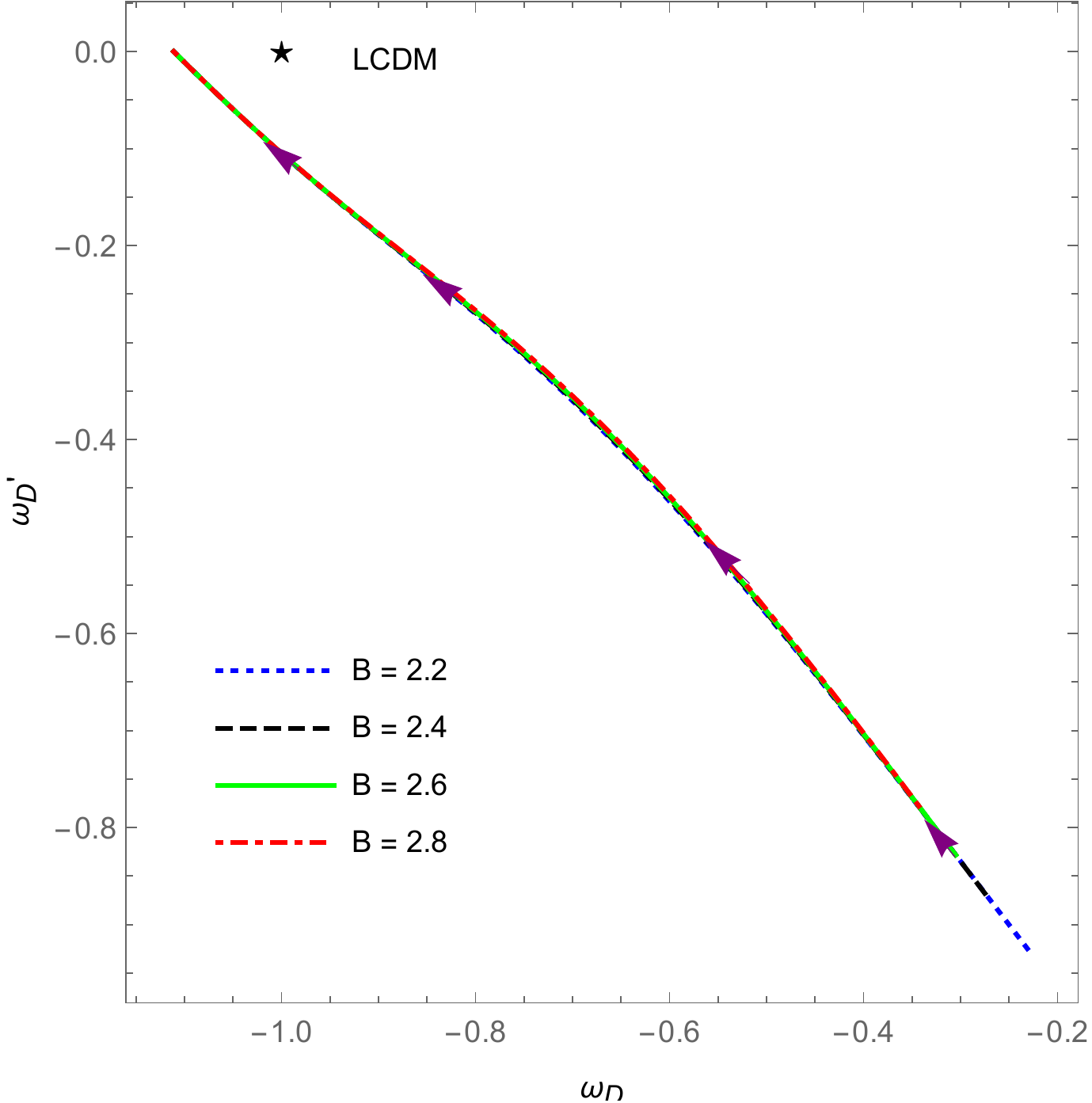}
			\caption{  The  diagram of $\omega_{D}-\omega_{D}^{'}$  for INTADE with initial conditions $\delta = 0.9$, $H(z=0)=67$,   $\Omega_{D}^{0} =0.73$, by considering $B=3$ and different values of coupling  $b^2$ (left panel) and different values of $B$ (right panel) fixing $b^2 = 0.1$.}
			\label{fig7}
			
		\end{center}
	\end{figure}
According to Caldwell and Linder \cite{ref78}, the  $\omega_{D}- \omega_{D}^{'}$ plane can be divided into two parts. One is for negative EoS parameter corresponding to positive behaviour of evolutionary parameter gives the thawing part $\omega_{D}< 0$, $\omega_{D}^{'} >0$, while for negative EoS parameter and negative behaviour of evolutionary parameter  $\omega_{D}< 0$, $\omega_{D}^{'} < 0$ get the freezing part of the evolving Universe.  
	The evolutionary trajectories of $\omega_{D}^{'}$ and $\omega_{D}$ plane are plotted in Fig. 7 for the INTDAE model for different coupling $b^2$ (left panel) and different values of  $B$ (right panel). This shows the INTDAE model lies in freezing region for all values of the coupling $b^2$. The evolutionary trajectories of the  $\omega_{D}- \omega_{D}^{'}$ plane are well-differentiated at all red shift region.  For $b^2 = 0$ the curve of $\omega_{D}- \omega_{D}^{'}$ pair lies in the quintessence $\omega >-1$ region and finally approaches the $\Lambda$CDM point $\omega_{D} = -1$, $\omega_{D}^{'}=0$  at late time.  While for other values of $b^2$	the  evolutionary trajectory of the  $\omega_{D}- \omega_{D}^{'}$ plane crosses the phantom divide line $\omega =-1$  and shows more deviation from the $\Lambda$CDM point $\omega_{D} = -1$, $\omega_{D}^{'}=0$ as the value of interaction increases. However,
		when the interaction between dark components is present, the situation becomes somewhat ambiguous
		because that the equation of state $\omega_{{D}}$ loses the ability of classifying dark energies definitely, due to the fact
		that the interaction makes dark matter and dark energy be entangled in each other. In this circumstance, the conceptions such as quintessence, phantom and quintom are not so clear as usual. But, anyway, we
		can still use these conceptions in an undemanding sense. It should be noted that when we refer to these
		conceptions the only thing of interest is the equation of state $\omega_{{D}}$ \cite{refu5}. The merit of the statefinder diagnosis method is that the
		statefinder parameters are constructed from the scale factor $a$ and its derivatives, and they are expected
		to be extracted in a model-independent way from observational data, although it seems hard to achieve
		this at present. While the advantage of the $(\omega_D,  \omega_D^{'})$
		analysis is that it is a direct dynamical diagnosis for dark energy \cite{ref78,ref79,refu6,refu7,refu8}. Therefore, the geometrical statefinder $(s-r)$ diagnosis and the dynamical $(\omega_D,  \omega_D^{'})$  diagnosis may be viewed as complementarity in some sense.

	\section{Conclusions}         
In this study, we have examined the INTADE model for different interactions ($b^2$) and parameter $B$ in a flat FLRW Universe by taking the conformal time as IR cutoff.  Statefinder and  $(\omega_D,\omega_D^{'})$ pairs have been used for diagnosing the INTADE model.  We can conclude our outcomes as:

\begin{itemize} 
	
	\item The scenario of interacting new Tsallis agegraphic dark energy leads
	to interesting cosmological phenomenology. The  evolutionary trajectories of  the deceleration parameter $q$ exhibits the transition from deceleration to acceleration happening  in agreement with observations.

	\item  	The EoS parameter of the INTADE model presents a rich behavior, and according to the strength of interaction   $b^2$  and value of the parameter $B$, it can be quintessence-like,
	phantom-like, or experience the phantom-divide crossing
	before or after the present time.\\
	
	\item  The first and second statefinder parameters $r(z)$ and $s(z)$ approach	to the $\Lambda$CDM model in the far future for the INTADE model and can be well-differentiated from the $\Lambda$CDM model at late time.\\
	
	\item   The evolutionary trajectory in $(s, r)$ plane of the INTADE model  approaches to the  $\Lambda$CDM point  ($r = 1$, $s = 0$) at late time. For  $b^2 = 0.2$,  $b^2 = 0.3$, the INTADE model  imitates the Chaplygin gas behaviour while for  $b^2 = 0.0$,  $b^2 = 0.1$, the INTADE model lies in the     quintessence region.\\  

		\item The curve of ($q, r$) plane of the INTADE model demonstrates that  it evolves near the
		matter dominated Universe at early time and approaches to the de Sitter expansion (SS) at the late time for different interactions ( $b^2$)   and  parameter $B$.\\

				\item The dynamical analysis $\omega_{D}- \omega_{D}^{'}$ pair, indicate that the INTADE model  lies in the freezing region for all values of the coupling $b^2$   and  parameter $B$.
\end{itemize}

	The statefinder analysis  has also been performed for the HDE, THDE, ADE and RDE models in literature as mentioned in the introduction.  The HDE, ADE  and THDE models have shown similar behaviour in the $ s - r $ plane to the INTADE model, while the INTADE model has different evolutionary behaviour as compared to the  RDE model in the $ s - r $ plane.\\
		
		In the future paper, we shall investigate other diagnostics such as Om diagnostics that can also be studied to understand the nature of the INTADE model.

\section*{Acknowledgments}
{  The author U. Sharma 
	thanks the IUCAA, Pune, India for awarding the visiting associateship.  We would like to thank the learned reviewer(s) for his valuable comments and suggestions that help us to improve this paper in the present form.}

\end{document}